# ONE-POT SYNTHESIS AND CHARACTERIZATION OF POLYYNES END-CAPPED BY BIPHENYL GROUPS (α,ω-BIPHENYLPOLYYNES)

Franco Cataldo[1(*)], Ornella Ursini[2], Alberto Milani[3], Carlo S. Casari[3]

[1]Actinium Chemical Research Institute, Via Casilina 1626 A, Rome, Italy
[2]CNR - Istituto di Metodologie Chimiche, Montelibretti, Roma
[3] Department of Energy, Politecnico di Milano via Ponzio 34/3, I-20133 Milano, Italy

**Abstract**

Stable polyyne chains terminated with biphenyl end groups (α,ω-biphenylpolyynes) were synthesized in a single step with an easy procedure by using the Cadiot-Chodkiewicz reaction conditions. The α,ω-biphenylpolyynes were separated through HPLC analysis and identified by means of their electronic absorption spectra. The α,ω-biphenylpolyynes were studied by FT-IR and Raman spectroscopy and the spectral interpretation was supported with DFT calculations. A peculiar stability of α,ω-biphenylpolyynes towards ozone was also observed.

## 1. Introduction

The sensational report about the production of a long polyyne chain of 6000 carbon atoms enclosed in a double wall carbon nanotube (DWCNT) [1] has re-awakened the enthusiasm about carbyne, the fabulous carbon allotrope made by an infinitely long sequence of carbon atoms with sp hybridization (with either alternated single and triple bonds or cumulated double bonds). Earlier attempts to the synthesis of carbyne are reviewed for example in the book of Heimann at al. [2].

*Corresponding author. Tel: 0039-06-94368230. E-mail: franco.cataldo@fastwebnet.it

The chemical approach toward the synthesis of this long acetylenic structure for example through the Glaser coupling reaction invariably led to a crosslinked carbonaceous solid [3,4] whose solid state [13]C-NMR analysis and Raman spectroscopy revealed indeed the presence of sp hybridized carbon atoms but mixed with sp$^2$ and sp$^3$ hybridized carbon atoms due to undesired crosslinking reactions [5,6]. A major issue in the development of sp-carbon systems is the highly unstable character of the acetylenic carbon chains [7,8]. For instance, sp-carbon films produced by cluster beams are stable only in vacuum, while the hydrogen-terminated polyynes and cyanopolyynes (polyynes terminated with nitrile groups) are stable only in dilute solutions and for limited time [9-15]. Some approaches to improve the stability of the polyynes chains have been proposed [16,17] and in the last few years many works have outlined that a feasible strategy to obtain stable carbon wires is based on the use of bulky end groups to end-cap the carbon chains [18-21]. Hence, the synthesis and the investigation of polyynes have seen a renewed interest as model systems towards carbyne [22,23] as well as complex alkyne-aromatic structures [24,25], which also led to the development of nanoscale systems with unique optical and electronic properties [26,27].

The chemical synthesis of polyynes involves several different approaches [9,19], ranging from the step synthesis involving oxidative coupling and the use of alkylsilane as protective end groups [28], to the laser ablation of carbon targets [29,30] as well as the submerged carbon arc [9-12] in which all polyynes are formed simultaneously in a single shot. It is indeed amazing that carbon arc and laser ablation conditions are mimicking the circumstellar environment [31-34] in which polyynes and cyanopolyynes are detected by radioastronomy and other techniques [35-37]. Polyynes are also present in the atmosphere of Titan the Saturn's giant moon, and their formation involves a complex photochemistry [38].

With regard to the synthesis of stable polyynes, one possibility is to adopt aryl groups as terminations. The Cadiot-Chodkiewicz coupling reaction [39] was proposed in combination with the use of diiodoacetylene as key synthon [40]. This synthetic approach was successfully employed in the one-pot synthesis of α,ω-diphenylpolyynes oligomers, which were separated and identified by HPLC-diode array [41]. Similarly, also α,ω-dinaphthylpolyynes oligomers were successfully synthesized combining the Cadiot-Chodkiewicz reaction with the use of diiodoacetylene [42]. The diphenylpolyynes were then studied by surface-enhanced Raman spectroscopy [30] and their non-linear optical properties were determined [43]. Optical, vibrational and electronic properties of these carbon-atom wires were investigated on polyynes [44] and different diarylacetylenes molecules [45-47] which are known to exhibit liquid crystals properties suitable for electro-optical applications [48-52].

Using the consolidated synthetic approach described above, the synthesis of biphenyl-terminated polyynes was successfully achieved and reported here. It should be noticed that biphenyl-terminated polyynes were already synthesized a few decades ago, but the synthetic path adopted at that time was based on a quite long and complex step synthesis [53]. On the other hand, in the present synthetic approach, where the Cadiot-Chodkiewicz reaction is combined with the use of diiodoacetylene, all the α,ω-biphenylpolyynes homologous series is obtained in a single reaction step. This simple synthetic approach leads to acetylenic chain length up to 12 sp-carbon atoms which are stable in air at room temperature even when the solvent is removed and a thin film is formed. Starting from easily accessible reagents, with no need of multiple step synthesis, no employment of troublesome reagents or of an inert atmosphere. the synthesis here reported is a significant step to make biphenyl-terminated polyynes easily accessible to everybody. The molecular wires so obtained were investigated by means of UV-VIS absorption spectroscopy and vibrational spectroscopy (Raman and FT-IR) to characterize the length and the structure of the chains. Experimental data are also supported by quantum chemical calculations for an accurate rationalization of the observed trends.

Biphenyl-terminated polyynes show a surprising resistance toward ozone attack, confirming that polyynes chains terminated with bulky end-groups are stable at ambient temperature even in absence of any solvent to prevent the crosslinking reactions. These findings gives further important information for the exploitation of carbon chain in future applications. Stability of sp-carbon structures is indeed a major issue since the typical argument against the development of sp-carbon systems is their high reactivity. The simple synthesis of stable wires is of great relevance in these grounds and in addition there is little knowledge about the reaction of ozone with polyynes in general. The exploration of the reaction of ozone with biphenyl-terminated polyynes demonstrates not only that these systems are stable to oxidation but contributes in shedding light onto the reaction of ozone with linearly conjugated systems.

**2. Experimental**

2.1 *Materials and equipment*

Diiodoacetylene was prepared as described elsewhere [40]. Ethynylbiphenyl was purchased from Sigma-Aldrich and used as received. All the other solvents and reagents used were analytical grades from Sigma-Aldrich. HPLC analysis was performed on Agilent Technologies 1100 apparatus equipped with an isocratic pump and a diode array detector. The HPLC column employed for the separation of the components was a common C8 column with a mobile phase composed by a mixture acetonitrile/water 80/20 or pure acetonitrile. With the former mixture the retention time of each biphenyl-substituted polyyne component was found too long. Instead, by using only acetonitrile the separation of the components was still effective in a considerable shorter time.

The electronic absorption spectra were recorded on a Shimadzu UV-2450 and the FT-IR spectra on a Nicolet 6700 from Thermo-Fisher. The FT-IR were recorded in transmittance mode on samples deposited from solutions on disposable polyethylene (PE) cards or on KBr disc.

The Raman spectra were obtained on a BWTEK spectrometer model BWS415i using a laser source at 785 nm. The spectra were acquired on the bis(biphenyl)polyynes solutions in decalin directly through a Pyrex vial.

2.2 *Synthesis of Cu(I) salt of ethynylbiphenyl*

Copper (I) salt of ethynylbiphenyl (copper biphenylacetylide) was prepared by dissolving ethynylbiphenyl (1.0 g) in 50 ml of ethanol and mixing the resulting solution with another solution of CuCl (1.0g) and hydroxylamine hydrochloride (0.5 g) in 30 ml of conc. ammonia solution. The resulting yellow precipitate was collected quantitatively through filtration using a common paper filter and an aspirator, washed thoroughly with water and ethanol and left to dry in air.

2.3 *Synthesis of the polyynes chains terminated with biphenyl groups (α,ω-biphenylpolyynes)*

Diiodoacetylene (5.4 g) was dissolved in 100 ml of decalin. The resulting solution was mechanically vigorously stirred with 1.3 g of copper biphenylacetylide in presence of 150 ml conc. $NH_3$, 40 ml of tetramethylenediamine (TMEDA) and 50 ml ethanol. The reaction mixture was stirred for 17 hours at room temperature. A gradual darkening of the mixture was noticed till a final black color. The reaction mixture was filtered through a paper filter "Rapida A" with the aid of an aspirator and the filtered solution was collected in two phases, aqueous and organic phases. The organic phase is composed almost exclusively by decalin, it was orange in color and was separated from the aqueous phase through a separatory funnel. The HPLC analysis of this crude reaction mixture has revealed that a series of undesired products were present in the mixture. More in detail, the presence of unreacted reagents were detected together with a series of polyynes half-terminated with biphenyl group from one side and an acetylene group on the other side.

2.4 *Purification of the crude reaction mixture*

To remove the unreacted reagents and the undesired polyynes with acetylene end-groups from the crude reaction mixture, the decalin solution was first shaken with an aqueous solution composed by CuCl (1.5 g), hydroxylamine hydrochloride (0.4 g), 50 ml of conc. ammonia solution and 20 ml of distilled water. A precipitate was slowly formed, left overnight and filtered out. The decalin solution was shaken with 200 ml HCl 10% to remove ammonia and TMEDA eventually present in the decalin solution which is now yellow-orange. After having discharged the HCl solution, decalin was washed two times with 300 ml each of distilled water inside a separatory funnel. After washing, the water was discharged. The final purification of the decalin solution was performed by shaking it with a solution of 1.0 g of $AgNO_3$ in 50 ml of conc. $NH_3$ and 50 ml of dist. water. A precipitate was observed. The mixture was stirred for a couple of hours, filtered through a paper filter and the collected decalin fraction filtered again through and Acrodisc syringe having 0.45 micron PVDF filter. The decalin solution now is yellow and the HPLC chromatogram shows exclusively the homologous series of biphenyl-terminated polyynes chains.

2.5 *HPLC analysis of purified solution*

The purified decalin (5 µL) was injected into the HPLC C8 column. Using only acetonitrile as mobile phase at a flow rate of 1.5 ml/min, the elution of 1,4-bis(biphenyl)-butadiyne (φ-φ-C≡C-C≡C-φ-φ) occurred at 1.651 min, followed by 1,6-bis(biphenyl)-hexatriyne (φ-φ-C≡C-C≡C-C≡C-φ-φ) at 1.784 min, 1,8-bis(biphenyl)-octatetriyne (φ-φ-C≡C-C≡C-C≡C-C≡C-φ-φ) at 2.086 min, 1,10-bis(biphenyl)-decapentiyne (φ-φ-C≡C-C≡C-C≡C-C≡C-C≡C-φ-φ) at 3.078 min, 1,12-bis(biphenyl)-dodecahexiyne (φ-φ-C≡C-C≡C-C≡C-C≡C-C≡C-C≡C-φ-φ) at 4.611 min. The symbol φ-φ- denotes the biphenyl group.

2.6 *Relative concentration of the α,ω-biphenylpolyynes*

The relative concentration of the α,ω-biphenylpolyynes having the general formula φ-φ -(C≡C)$_n$-φ-φ was determined from the electronic absorption spectra measured with the diode array detector. For the oligoyne n=2 the relative concentration in the mixture was 77.1% by mol, followed by n=3 at a concentration of 18.4%, n=4 at a relative concentration of 2.6%, n=5 at a relative concentration of 1.7% and finally, n=6 at a relative concentration of 0.2%.

2.7 *Synthesis of the 1,4-bis(biphenyl)-butadiyne (φ-φ -C≡C-C≡C-φ-φ) to be used as standard*

In order to recognize the first member of the homologous series of α,ω-biphenylpolyynes, 1,4-bis(biphenyl)-butadiyne (φ-φ-C≡C-C≡C-φ-φ) was synthesized by direct coupling reaction of ethynylbiphenyl under the Eglington conditions [52]. Ethynylbiphenyl (96 mg) was dissolved in 40 ml of ethanol and treated with 55 mg of CuCl and 88 mg of Cu(II) acetate. The coupling reaction starts when pyridine (30 ml) was added to the reaction mixture at room temperature in presence of aqueous ammonia. The reaction mixture was stirred at room temperature for 12 hours. Afterwards, the blue reaction mixture was treated with 10% HCl until it became completely green. The reaction mixture was then extracted with 35 ml of n-hexane and bis(biphenyl)-butadiyne (φ-φ-C≡C-C≡C-φ-φ), the unique product used to record the electronic absorption spectrum and the retention time in the HPLC column.

2.8 *Ozonation of α,ω-biphenylpolyynes*

A decalin solution (20 ml) of α,ω-biphenylpolyynes mixture having a total concentration of 0.4 mM, was exposed to a stream of ozonized oxygen in a Drechsel gas washing apparatus. The unreacted ozone passed through the decalin solution was collected in another gas washing bottle charged with a 5% KI aqueous solution. It was immediately evident that only a fraction of ozone passing in the decalin solution was reacting with the polyynes. However the largest ozone fraction was collected in the KI solution.

2.9 *DFT calculations*

Geometry optimization, IR spectra calculations and the prediction of the electronic absorption spectra have been computed by means of Density Functional Theory (DFT) and Time-dependent DFT (TDDFT). The calculations were carried out for the isolated molecule model reported in Fig. 1b for all the different chain lengths from n = 1 to n = 6. The hybrid PBE0 exchange-correlation functional and the cc-pVTZ basis set have been used and the calculations have been carried out with the Gaussian09 quantum chemistry code [Gaussian 09, Revision A.02, M. J. Frisch, G. W. Trucks, H. B. Schlegel, G. E. Scuseria, M. A. Robb, J. R. Cheeseman, G. Scalmani, V. Barone, G. A. Petersson, H. Nakatsuji, X. Li, M. Caricato, A. Marenich, J. Bloino, B. G. Janesko, R. Gomperts, B. Mennucci, H. P. Hratchian, J. V. Ortiz, A. F. Izmaylov, J. L. Sonnenberg, D. Williams-Young, F. Ding, F. Lipparini, F. Egidi, J. Goings, B. Peng, A. Petrone, T. Henderson, D. Ranasinghe, V. G. Zakrzewski, J. Gao, N. Rega, G. Zheng, W. Liang, M. Hada, M. Ehara, K. Toyota, R. Fukuda, J. Hasegawa, M. Ishida, T. Nakajima, Y. Honda, O. Kitao, H. Nakai, T. Vreven, K. Throssell, J. A. Montgomery, Jr., J. E. Peralta, F. Ogliaro, M. Bearpark, J. J. Heyd, E. Brothers, K. N. Kudin, V. N. Staroverov, T. Keith, R. Kobayashi, J. Normand, K. Raghavachari, A. Rendell, J. C. Burant, S. S. Iyengar, J. Tomasi, M. Cossi, J. M. Millam, M. Klene, C. Adamo, R. Cammi, J. W. Ochterski, R. L. Martin, K. Morokuma, O. Farkas, J. B. Foresman, and D. J. Fox, Gaussian, Inc., Wallingford CT, 2016]. Since DFT methods are known to overestimate π-electron conjugation, when comparing vibrational spectra with the experimental ones, scaling procedures are needed and, due to similarity with diphenylpolyynes investigated in the past [41-47], we adopted the same chain length-dependent scaling factors also in the current case. These scaling factors are 0.9503, 0.9487, 0.9508, 0.9522 and 0.9565 for n=2, n=3, n=4, n=5 and n=6 respectively. For n=1 the scaling factor was not evaluated and here we used the same factor as in n=2.

## 3. Results and discussion

3.1 *Chemical structure of α,ω-biphenylpolyynes series from DFT simulation*

The chemical structure of the α,ω-biphenylpolyynes series is shown in Fig. 1 together with the sketch of the optimized geometry obtained by DFT simulation. The main feature of this class of compounds regards the presence of the biphenyl end groups which is characterized by two benzene rings connected each other by a rigid single bond. In the solid state the two rings are coplanar but in solution and in the gas phase the two rings are expected to be rotated each other by an angle of 45° due to sterical repulsions [54].

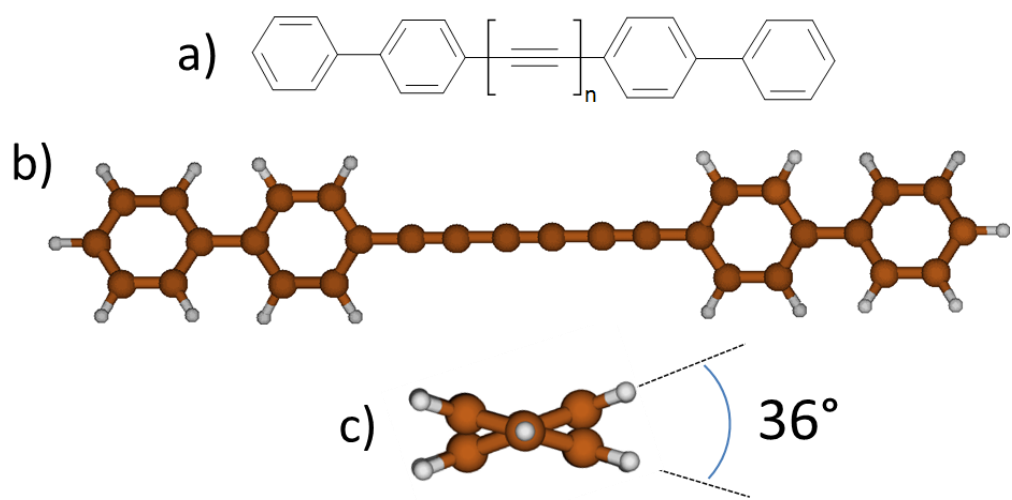

Fig. 1 – a) Chemical structure of the α,ω-biphenylpolyynes series; b) DFT calculated optimized geometry of 1,6-bis(biphenyl)-hexatriyne in top view and c) side view showing the DFT computed torsional angle between phenyl groups.

This is indeed supported by DFT optimized geometries where a torsional angle of 36° degrees is predicted between the two rings, as sketched in Fig. 1c. In the solid state the sterical repulsions forces are compensated by the lattice forces which of course are not present in solution or in the gas phase

[54]. On the other hand the polyyne chain connecting the end groups can be viewed as a rigid linear chain. Only for very long chains (e.g n > 10) the polyyne chain in the solid state can present some deviations from linearity [19,20].

3.2 *Observations on the synthesis of α,ω-biphenylpolyynes*

The use of the Cadiot-Chodkiewicz synthetic reaction conditions [55] combined with the use of diiodoacetylene and copper(I)biphenylacetylide as reagents implies that together with the desired α,ω-biphenylpolyynes also half-terminated polyynes are formed, (i.e. φ-φ-(C≡C)$_n$-H) as by-products. Therefore, it was necessary to remove such undesired products. Fortunately, these polyynes can be removed by precipitation with the addition of Cu(I) and Ag(I) salts. Indeed, the purification step described in detail in the experimental section has involved the treatment of the reaction mixture first with Cu(I) salt and later with Ag(I) to ensure the complete removal of the undesired by-products.

Fig. 2 shows the electronic absorption spectrum of the α,ω-biphenylpolyynes mixtures, after the purification step. The absorption bands at 295, 310, 330 and 354 nm are assignable to to 1,4-bis(biphenyl)-butadiyne (φ-φ-C≡C-C≡C-φ-φ), which was prepared with the Eglington coupling reaction [55] (see the experimental section) purposely to use it as reference for the UV-VIS and for the retention time in the HPLC column.

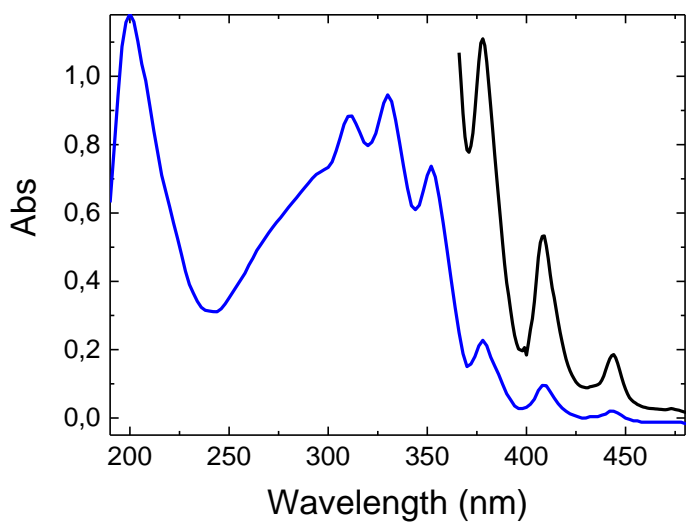

Fig.2 – Electronic absorption spectrum of α,ω-biphenylpolyynes mixture fully purified from by products. The absorption bands at shorter wavelengths are due to 1,4-bis(biphenyl)-butadiyne (φ-φ-

C≡C-C≡C-φ-φ), those at longer wavelengths are due to 1,6-bis(biphenyl)-hexatriyne (φ-φ-C≡C-C≡C-C≡C-φ-φ) and higher homologues. The black curve is a magnification (x5) of the blue curve. (A colour version of this figure can be viewed online.)

Our synthesis approach presents the advantage that polyynes of this type are produced all together in a single shot. The separation and identification of each component of the series can be achieved by HPLC analysis using the diode array detector to record the electronic absorption spectrum of each component, a procedure which was successfully employed in the separation of polyynes produced with the carbon arc [9-12] and with phenyl and naphthyl end groups [41,42]. In these previous works the HPLC analysis was performed under isocratic conditions using a mobile phase composed by acetonitrile/water 80:20. With this mobile phase the retention time of the α,ω-biphenylpolyynes became extremely long, so that the higher members of the series are eluting by far after 100 min. Furthermore the elution peaks were extremely broadened especially for the higher members of the series. The solution to this problem was to use exclusively acetonitrile as mobile phase. In this way the retention time of all the α,ω-biphenylpolyynes was deeply reduced although the peak separation remained sufficiently high to permit a clear separation of the various components of the series.

3.3 *Electronic absorption spectra of the α,ω-biphenylpolyynes*

The HPLC separation of the α,ω-biphenylpolyynes series permitted to record the individual electronic absorption spectrum of each member of the series which are reported in Fig. 3 and 4.
In Fig. 3 the electronic absorption spectrum of ethynylbiphenyl is shown as a reference although its retention time was studied with the C8 HPLC column together with a standard sample of 1,4-bis(biphenyl)-butadiyne (φ-φ-C≡C-C≡C-φ-φ) which was prepared by a separate synthesis. Knowing the retention times of the reagent ethynylbiphenyl and of the first member of the α,ω-biphenylpolyynes series, has permitted a straightforward assignment of the retention times and the associated electronic absorption spectra of the higher homologues of the series. All the oligomers were identified starting from n = 2 (see Scheme 1) to n = 6 and the longest homologue detected was composed by 12 acetylenic carbon atoms.
Fig. 4 shows a comparison between the position of the longest wavelength transition of the α,ω-diphenylpolyynes series studied previously [41] and the current α,ω-biphenylpolyynes series. As reported in the graph, the longest wavelength transition of the electronic absorption spectra is shifted linearly toward longer wavelength by the addition of each C≡C unit. The slope results slightly more steeply for the diphenylpolyynes series rather than the α,ω-biphenylpolyynes series. In fact, the two lines crosses each other at n = 6.

$$\lambda_{diphenylpolyynes} = 34.6\,n + 257.9 \qquad (1)$$

$$\lambda_{\alpha,\omega\text{-biphenylpolyynes}} = 27.2\,n + 301.6 \qquad (2)$$

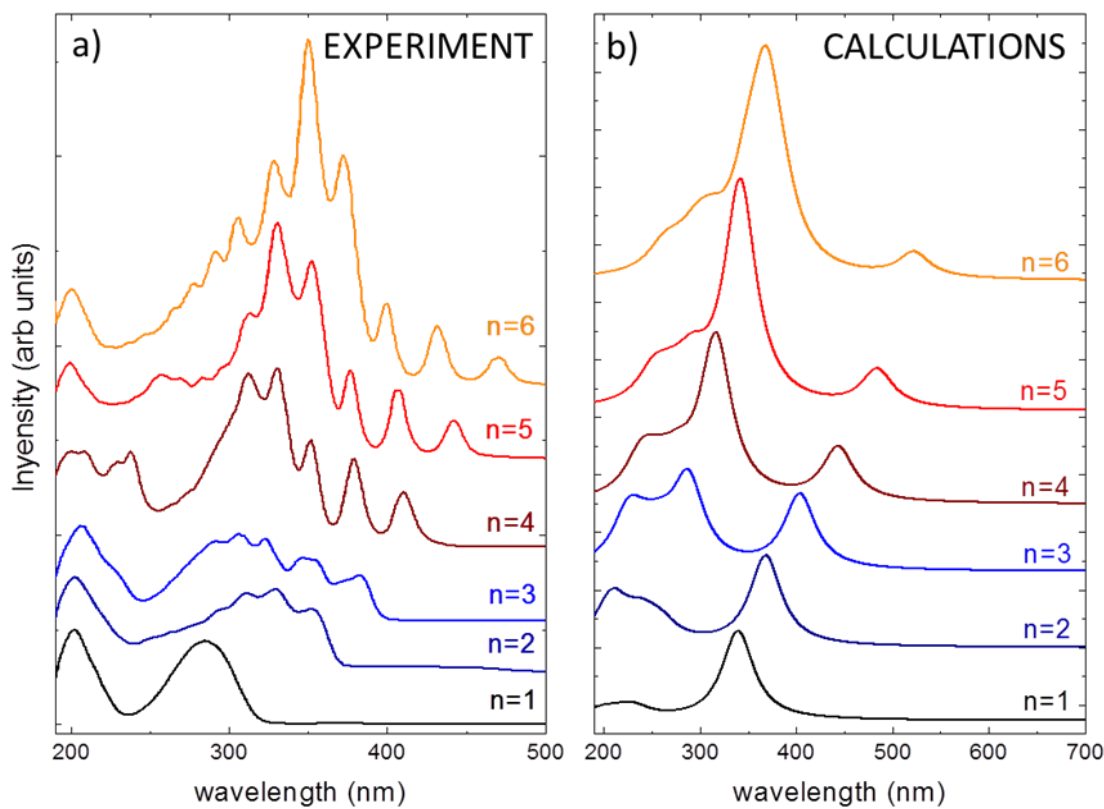

Fig. 3 - a) Electronic absorption spectra as registered by the diode array detector: ethynylbiphenyl, 1,4-bis(biphenyl)-butadiyne (φ-φ-C≡C-C≡C-φ-φ) with Rt=1.651 min, followed by 1,6-bis(biphenyl)-hexatriyne (φ-φ-C≡C-C≡C-C≡C-φ-φ) with Rt =1.784 min, 1,8-bis(biphenyl)-octatetriyne (φ-φ-C≡C-C≡C-C≡C-C≡C-φ-φ) with Rt=2.086 min, 1,10-bis(biphenyl)-decapentiyne (φ-φ-C≡C-C≡C-C≡C-C≡C-C≡C-φ-φ) with Rt=3.078 min, 1,12-bis(biphenyl)-dodecahexiyne (φ-φ-C≡C-C≡C-C≡C-C≡C-C≡C-C≡C-φ-φ) with Rt=4.611 min. b) Absorption spectra calculated by means of TDDFT computations.

It is also interesting to note that the intercept of eq. 1 for the diphenylpolyynes series is found at about 258 nm and when n = 0 the series converges into a biphenyl whose longest wavelength transition is found at 247 nm [56], not far from the 257.9 nm suggested by eq.1. Instead the intercept for the α,ω-biphenylpolyynes series converges on the quaterphenyl whose longest wavelength transition occurs at 294 nm [56] again not far from the 301.6 nm suggested by eq.2.

For n = 1 tolane (diphenylacetylene) is the first member diphenylpolyynes series with the longest wavelength transition at 295 nm [56] which is in fair agreement with the value 292.5 nm calculated from eq. 1. On the other hand, for n = 1 in the case of α,ω-biphenylpolyynes series, we are dealing with bis(biphenyl)acetylene whose longest wavelength transition is expected at 328.8 nm. The calculated Δλ between the two series is 328.8 - 292.5 = 36.3 nm.

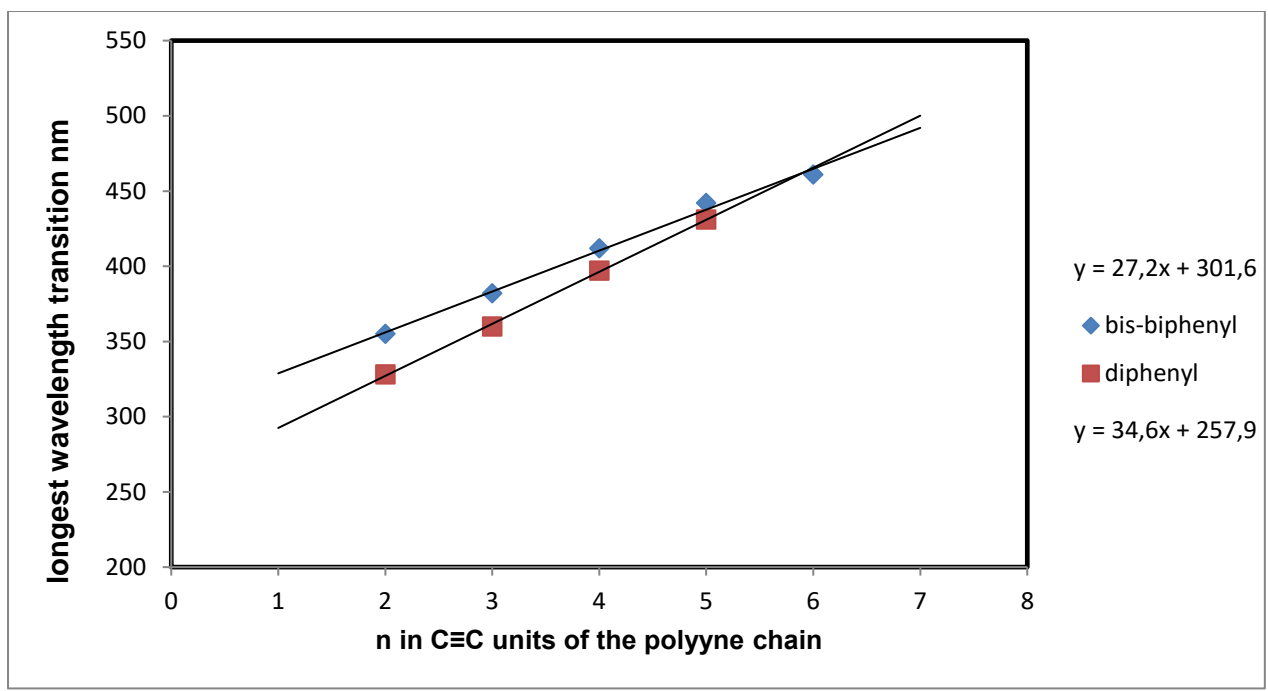

Fig. 4 – Position of the longest wavelength transition of the two homologous series α,ω-diphenylpolyynes series studied previously [41,43] and the current α,ω-biphenylpolyynes series.

Other authors have studied the electronic absorption spectra of α,ω-biphenylpolyynes series in tetrahydrofuran (THF) instead of decalin used in the present work [53]. The best fitting of the longest wavelength transition was achieved through the following equation [53]:

$$\lambda_{\alpha,\omega\text{-biphenylpolyynes}} = 24.0\ n^{1.1} + 304 \qquad (3)$$

which is quite similar to eq. 2 found in decalin. According to the authors of ref. [53], the exponent of n is 1.0 for the diphenylpolyynes series and changes to 1.1 for bulkier biphenyl end groups reaching 1.5 for the α,ω-diflorenylpolyyne series.

In Fig. 3b, the TDDFT computed absorption spectra are reported for the different chain lengths. In agreement with the experimental spectra, a peculiar shift of the bands to longer wavelengths is observed for increasing chain lengths as a results of the increasing pi-electron conjugation. In the computed spectra the detailed vibronic transitions observed in the experiments have been not computed and all the bands are found at a larger wavelength as a result of the overestimation of conjugation of DFT. A general agreement with the experimental spectra is observed however in the pattern evolution: in both cases it is indeed well-evident the peculiar evolution from shorter to longer chains. In shorter chains, the dominant contributions are found at higher wavelength (computed at about 380 nm for n=2 and shifting up to 520 nm for n=6). This transition is indeed mostly described as the HOMO-LUMO excitation and, as expected, it shows an upward shift in wavelength with chain length. In both the experimental and theoretical spectra, this band becomes less and less intense with respect to another band found at lower wavelengths and which becomes the dominant one starting from n=4. This band is associated to many different contributions in terms of one-electron excitations, a detailed computational analysis would be required for a complete rationalization, but this is beyond the aims of this paper.

Vibrational spectroscopy was adopted as well assessed technique to investigate sp-carbon systems [43-45].

3.4 *Vibrational spectroscopy of the α,ω-biphenylpolyynes*

As already discussed in the previous sections, the purified α,ω-biphenylpolyynes series was prepared in decalin solvent. To record the FT-IR spectra, the α,ω-biphenylpolyynes solution in decalin was deposited on a disposable PE card and on a KBr disc and the solvent was left to evaporate. The FT-IR spectra on the evaporation residues, corresponding to the α,ω-biphenylpolyynes mixture are shown in Fig. 5.

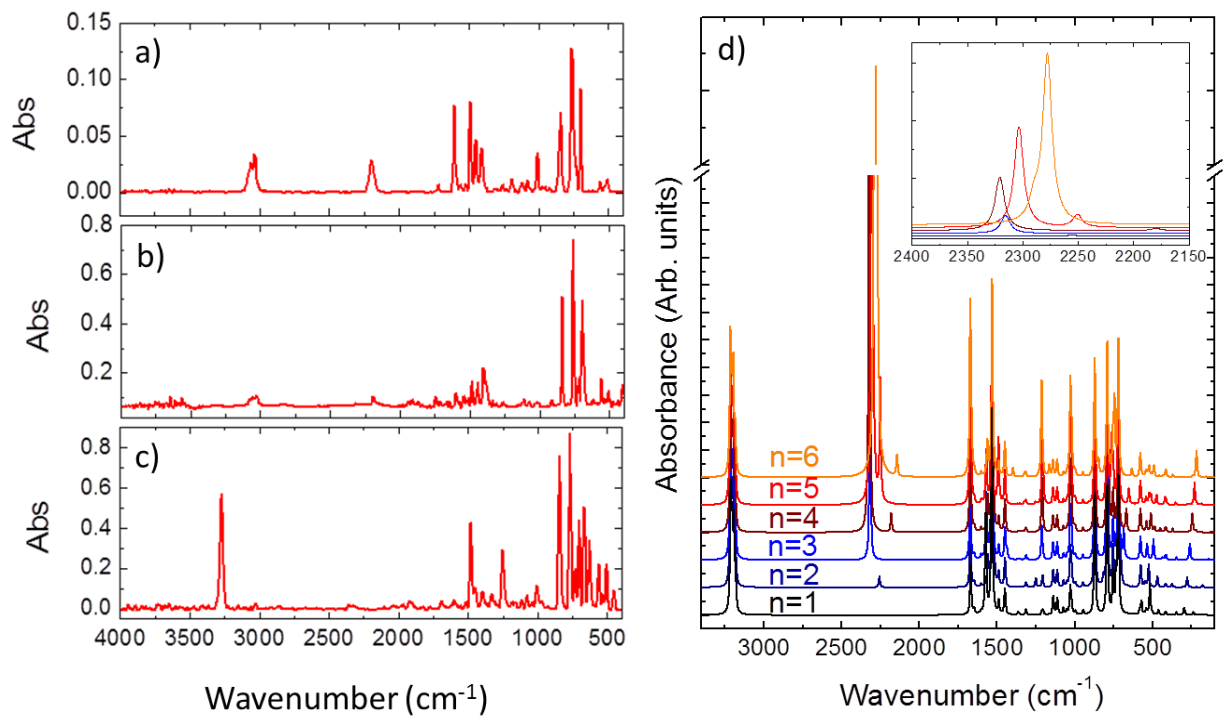

Fig. 5 – FT-IR spectra: a) Bis(biphenyl)polyynes deposited on PE card; b) bis(biphenyl)polyynes deposited on KBr disc; c) ethynylbiphenyl monomer in KBr pellet. d) DFT computed IR spectra (scaled frequency, see text)

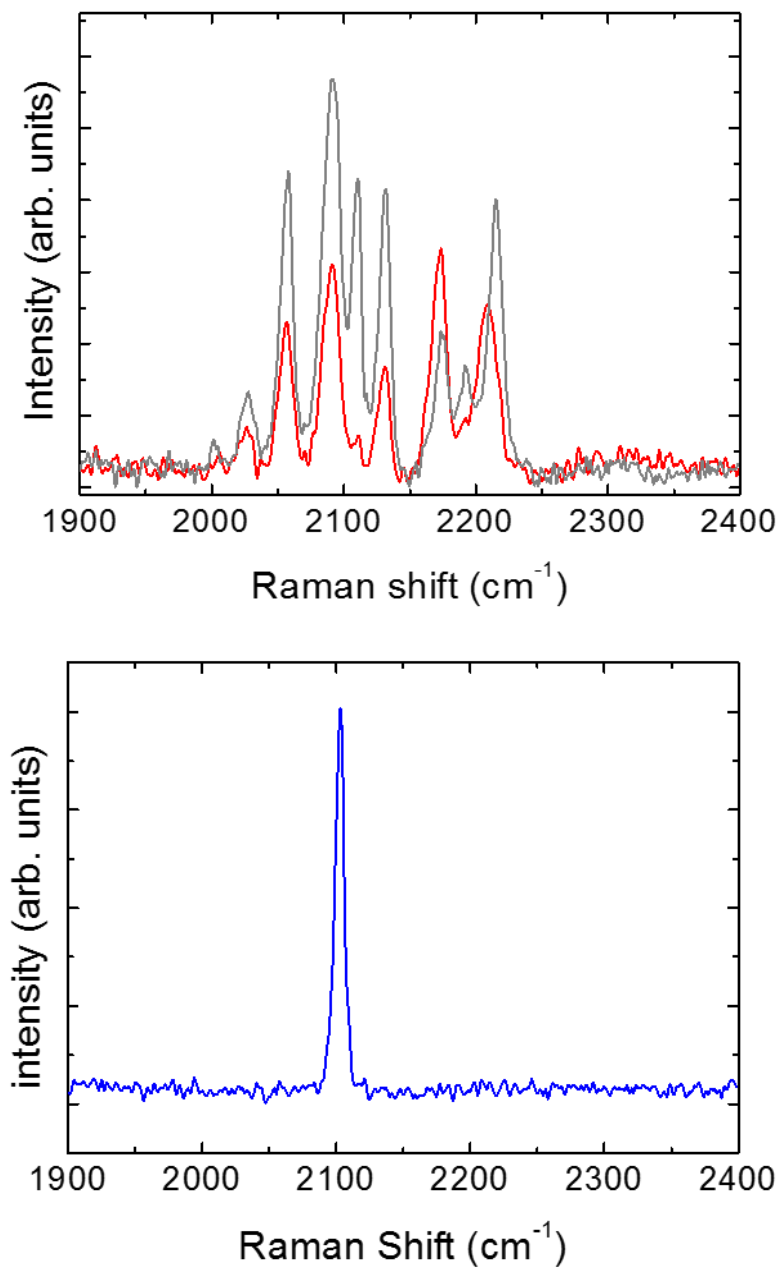

Fig. 6 – Raman spectra. (Top figure) Crude α,ω-biphenylpolyynes series (red trace); α,ω-biphenylpolyynes series after purification with $Cu^+$ and $Ag^+$ treatment (blue trace). The crude spectrum is characterized by the following bands 2028; 2058; 2091 ; 2111; 2131; 2173; 2192; 2216 $cm^{-1}$. After the removal of terminal acetylenes like ethynylbiphenyl (see spectrum in red), the band at 2111 $cm^{-1}$ is removed together with the band at 2192 $cm^{-1}$. The band originally at 2216 $cm^{-1}$ appears shifted toward lower frequencies: 2208 $cm^{-1}$. (bottom Figure, blue trace) Raman spectrum of ethynylbiphenyl, characterized by a sharp band at 2103 $cm^{-1}$.

There are little differences between the two spectra obtained on PE card on on KBr, although the former appears better resolved. Fig. 5 also shows the FT-IR spectrum of ethynylbiphenyl, the starting monomer from which the α,ω-biphenylpolyynes mixture were synthesized. The infrared spectrum of ethynylbiphenyl is characterized by a strong ≡C-H stretching band at 3274 cm$^{-1}$, and by the ≡C-H bending modes at 625 and 664 cm$^{-1}$ [57]. These bands are completely absent in the FT-IR spectra of the α,ω-biphenylpolyynes mixture shown in Fig. 5, confirming its complete reaction to the desired polyynes or removal of the unreacted ethynylbiphenyl. The α,ω-biphenylpolyynes mixture FT-IR spectra are instead characterized by a medium-weak triple bond stretching band at 2196 cm$^{-1}$. The weakness of this band is typical for symmetric acetylenes [57]. Other characteristic bands of the α,ω-biphenylpolyynes mixture are the aromatic C-H stretching bands appearing at 3078, 3058 and 3030 cm$^{-1}$ obviously due to the biphenyl end groups of the polyynes chains. The DFT computed IR spectra are reported in Fig. 5d and support the band assignment discussed above.

In parallel with FT-IR, Raman spectroscopy is the best analytical technique for the detection of the conjugated C≡C triple bond stretching in acetylenes [6]. Raman spectroscopy is so versatile that it is possible to follow the end-capped polyynes formation practically in situ, by collecting Raman spectra on the decalin phase from the external of the reaction flask. Simply, the laser beam of the Raman spectrometer is placed on the external walls of the reaction flask. Fig. 6 shows a Raman spectrum in the C≡C stretching region of the reaction mixture in decalin. The slow decay of the intensity of the band at 2111 cm$^{-1}$ associated to ethynylbiphenyl (see Fig. 6 at bottom, for comparison), corresponds to the growth of all the other bands due to the formation of the α,ω-biphenylpolyynes mixture. In the selected reaction conditions, the formation of α,ω-biphenylpolyynes mixture is indeed a slow reaction. Thus, at a certain point, the excess of unreacted ethynylbiphenyl and other by-products with terminal acetylene groups were removed by shaking the decalin solution with Cu$^+$ and Ag$^+$ solutions. Fig. 6 shows also the Raman spectrum of the purified solution: the band at 2111 cm$^{-1}$ due to residual ethynylbiphenyl is removed together with the band at 2192 cm$^{-1}$ due to a by-product with terminal acetylene group. The series of Raman bands observed in the spectrum of Fig. 6 of the purified reaction mixture are due to all the α,ω-biphenylpolyynes. As reported in great detail in another work [58], the band at 2208 cm$^{-1}$ was assigned to 1,4-bis(biphenyl)-butadiyne (φ-φ-C≡C-C≡C-φ-φ), followed by 1,6-bis(biphenyl)-hexatriyne (φ-φ-C≡C-C≡C-C≡C-φ-φ) with the Raman band at 2173 cm$^{-1}$, 1,8-bis(biphenyl)-octatetriyne (φ-φ-C≡C-C≡C-C≡C-C≡C-φ-φ) with the Raman band at 2131 cm$^{-1}$, 1,10-bis(biphenyl)-decapentiyne (φ-φ-C≡C-C≡C-C≡C-C≡C-C≡C-φ-φ) with the Raman band at 2091 cm$^{-1}$ and 1,12-bis(biphenyl)-dodecahexiyne (φ-φ-C≡C-C≡C-C≡C-C≡C-C≡C-C≡C-φ-φ) with the Raman band at 2058 cm$^{-1}$ (As usual, the symbol φ-φ- denotes the biphenyl group). A detailed discussion of the Raman spectra based on DFT calculations, also referring to other similar systems, has been presented elsewhere [58].

3.5 *Ozonation of α,ω-biphenylpolyynes in decalin*

Following the structural characterization of the synthesized chains, we investigated their behavior during ozonations. The ozonation of acetylene and simple alkynes has received a certain attention [59,60] also in recent times [61-63], but little work was instead dedicated to the ozonation of oligoynes and polyynes [13,64,65]. More in detail, the ozonation of hydrogen-terminated polyynes yields an oxidized polymeric product which precipitates out from the reaction medium [13,63]. The precipitate is a polyozonide, formed by a series of crosslinking reactions of the ozonized polyynes. Since the ozonized polyynes are removed out from the reaction medium by precipitation, they still maintain unreacted acetylenic bonds as revealed by FT-IR spectroscopy of the product [13]. It was shown [65], that in a competitive ozonolysis of a mixture of polyynes and cyanopolyynes, the ozone attack is directed selectively toward the polyynes because the cyano substituent in a polyyne chain exerts a deactivating effect due to its electron withdrawing action which reduces the charge density in the triple bonds through a conjugation effect along the polyacetylene chains. Biphenyl end groups in the α,ω-biphenylpolyynes exert also a moderate electron withdrawing action reducing the reactivity of ozone with the polyacetylene chains, although less than in the case of the cyano end-groups.

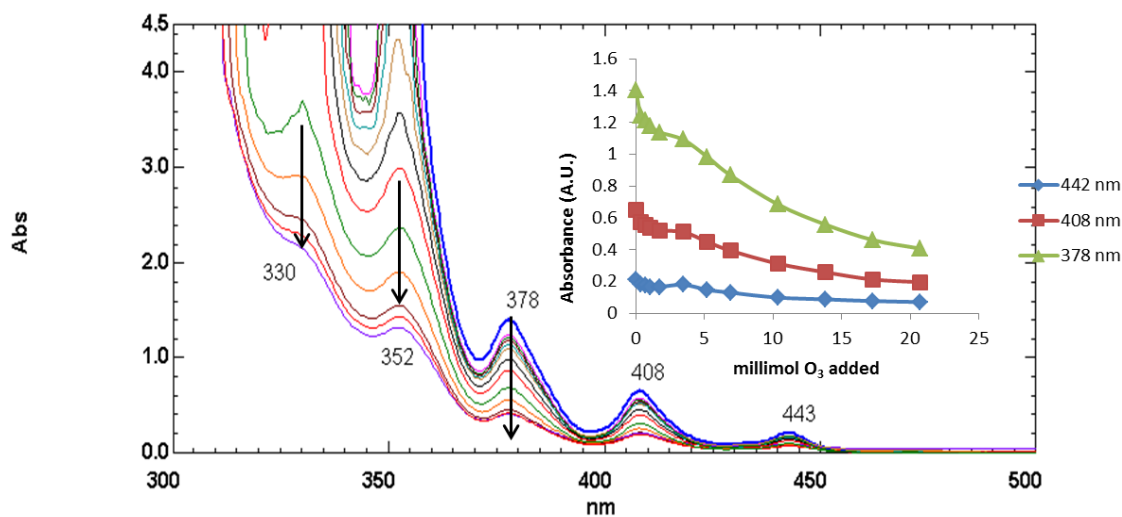

Fig. 7 – Electronic absorption spectra of α,ω-biphenylpolyynes mixture treated with ozone. All the absorption bands are gradually reduced in intensity because of the decomposition of the polyynes into other oxidized derivatives. The arrows indicate the trend of all the bands as function of the ozone added. Trend of the absorption bands due to bis(biphenyl)-octatetriyne (φ-φ-C≡C-C≡C-C≡C-C≡C-φ-φ) at 378 nm, 1,10-bis(biphenyl)-decapentiyne (φ-φ-C≡C-C≡C-C≡C-C≡C-C≡C-φ-φ) at 408 nm, 1,12-bis(biphenyl)-dodecahexiyne (φ-φ-C≡C-C≡C-C≡C-C≡C-C≡C-C≡C-φ-φ) at 442 nm as function of the ozone added to the solution are shown in the inset.

Fig. 7 shows the changes introduced by the action of the ozone on the α,ω-biphenylpolyynes: a gradual decay of the intensity of the absorption bands it is observed due to the oxidation and decomposition of the polyyne chain. The data of Fig. 7 are reported in the inset of Fig. 7 for some α,ω-biphenylpolyynes. The most important observation here is that the reactivity of α,ω-biphenylpolyynes with ozone is really small, since large amounts of ozone were passed into the solution and only a small fraction of ozone reacted with the α,ω-biphenylpolyynes. Normally with alkenes the reaction efficiency is 100% [66], but with α,ω-biphenylpolyynes the reaction with ozone is so slow that large part of it passed through the solution unchanged. The other interesting aspect is the complete absence of polyozonides, the precipitate observed in the ozonation of hydrogen-terminated polyynes. Evidently, the crosslinking and polymerization of polyynes with ozone occurs due to the ozone attack of the hydrogen end groups other than the triple bonds. In the case of α,ω-biphenylpolyynes, the bulky biphenyl end groups hinder completely the oxidative polymerization reaction and the ozone attack is directed solely toward the conjugated triple bonds. This behavior is peculiar and different with respect to other systems based on linear sp-carbon. For instance thin films of disordered sp-$sp^2$ network produced by supersonic beams of carbon clusters in vacuum show a high reactivity with oxygen which induces transformation of sp carbon into $sp^2$ [7,67]. Moving closer to isolated linear carbon chains it was shown that H-terminated polyynes in liquid produced by the submerged arc discharge are prone to degradation due to cross linking reactions [15]. The use of large terminating groups to improve stability in sp-carbon chains is quite recent and appears promising to extend the field of sp-carbon from fundamental science to applications. With this aim the detailed investigation of stability even against ozonation is fundamental to understand the mechanisms responsible of degradation.

## 4. Conclusions

α,ω-Biphenylpolyynes mixture was successfully synthesized using the Cadiot-Chodkiewicz reaction conditions using diiodoacetylene and Cu(I) biphenylacetylide. The α,ω-biphenylpolyynes mixture was successfully separated into its components by HPLC analysis and the electronic absorption spectrum of each component was recorded and studied. The complete series of α,ω-biphenylpolyynes from n=2 to n=6 was produced (see Fig. 1) and identified through the electronic absorption spectra and Raman spectroscopy. The evidence of the successful α,ω-biphenylpolyynes was also obtained from the FT-IR analysis and from the Raman spectroscopy supported by DFT calculations. The reactivlty of α,ω-biphenylpolyynes toward ozone was assayed in a preliminary way and it was found a surprisingly low reactivity. This fact is certainly due to the electron withdrawing effects of the biphenyl end groups which reduce the charge density in the conjugated triple bonds, partly inhibiting the electrophilic ozone attack. The overall stability shown both when the systems are deposited on a substrate with the solvent left drying and toward exposition to ozone indicates that highly stable polyyne-lke systems can be produced and makes carbon wires a reliable nanosystem for fundamental and applied materials science.